\documentclass[aip,jap,reprint]{revtex4-1}
\usepackage{graphicx}
\usepackage{dcolumn}
\usepackage{bm}
\usepackage{url}

\usepackage[colorlinks=true,linkcolor=blue]{hyperref}
\expandafter\ifx\csname package@font\endcsname\relax\else
\expandafter\expandafter
\expandafter\usepackage
\expandafter\expandafter
\expandafter{\csname package@font\endcsname
\fi
\hyphenation{title}

\begin{document}

\title{Search for plant biomagnetism with a sensitive atomic magnetometer}

\author{Eric Corsini}
\email{eric.corsini@gmail.com}
\affiliation{Department of Physics, University of California, Berkeley, CA 94720-7300}
\author{Victor Acosta}
\affiliation{Department of Physics, University of California, Berkeley, CA 94720-7300}
\author{Nicolas Baddour}
\affiliation{Department of Physics, University of California, Berkeley, CA 94720-7300}
\author {James Higbie}
\affiliation{Bucknell university, 701 Moore Avenue, Lewisburg, PA 17837}
\author{Brian Lester}%
\affiliation{Department of Physics, California Institute of Technology, Pasadena, CA 91125}
\author{Paul Licht}
\affiliation{UC Botanical Garden, University of California, Berkeley, CA 94720}
\author{Brian Patton}
\affiliation{Department of Physics, University of California, Berkeley, CA 94720-7300}
\author{Mark Prouty}
\affiliation{Geometrics Inc., 2190 Fortune Drive, San Jose, CA 95131}
\author{Dmitry Budker}
\homepage{http://budker.berkeley.edu/}
\affiliation{Department of Physics, University of California, Berkeley, CA 94720-7300}
\affiliation{Nuclear Science Division, Lawrence Berkeley National Laboratory, Berkeley CA 94720}

\date{\today}

\begin{abstract}
We report what we believe is the first experimental limit placed on plant biomagnetism.  Measurements with a sensitive atomic magnetometer were performed on the Titan arum \textit{(Amorphophallus titanum)} inflorescence, known for its fast bio-chemical processes while blooming. We find that the magnetic field from these processes, projected along the Earth's magnetic field, and measured at the surface of the plant, is~$\lesssim$~0.6\nobreak{ }$\mu$G.
\end{abstract}
\maketitle

\section{\label{intro}Introduction}
With the advent of sensitive magnetometers, the detection of biological magnetic signals (pioneered in the 1960s \cite{early}) has added a new dimension to the understanding of physiological and biological processes by providing more information about the source of the associated electrical currents than surface electrodes\cite{IEEE-Cohen,MEGvEEG-Cohen}.  Sensitive magnetic field measurements have enabled advances in magnetoencephalography, magnetoneurography, and magnetocardiography\cite{Cohen,Wiley,weiss}. Magnetic fields from the heart, the result of cardiac action potential with electrical current densities that can reach $\rm \sim 100 ~Am^{-2}~$, are on the order of 1~$\rm\mu$G, when measured at or near the skin surface.   Another example is the measurement of magnetic fields associated with human brain functions, of the order of 1~nG, which has given a new understanding in the organization of neural systems underlying memory, language, and perception, as well as the diagnosis of related disorders\cite{NIH-MEG}.
\begin{figure}
\centering
  \includegraphics[width=2.5in]{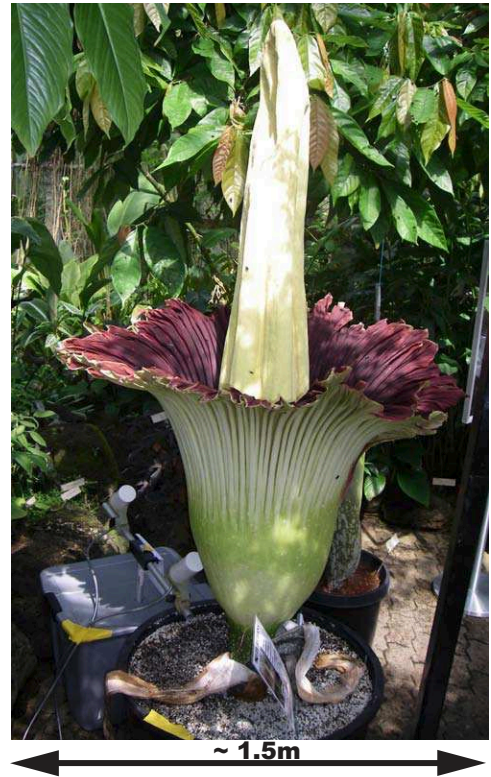}
  \caption{The titan arum (or \textit{Amorphophallus titanum}), nicknamed `Trudy',  in full bloom on June~23,~2009, at the University of California Botanical Garden. The Geometrics G858 magnetometer sensors are visible behind the plant on the left.}
\label{fig:titan}
\end{figure}

Superconducting quantum interference device (SQUID) magnetometers have been leading the field of ultra-sensitive magnetic field measurements since the 1960s\cite{Clarke1,Clarke2}.  However, resonant magneto-optics and atomic magnetometry\cite{nature} have experienced a resurgence driven by technological developments, specifically by the advent of reliable, inexpensive, and easily tunable diode lasers, and by refinements of the techniques for producing dense atomic vapors with long-lived polarized ground-states. These advances have enabled atomic magnetometers to achieve sensitivities rivaling and even surpassing that of the SQUID, and have a dynamic range from near-zero field to Earth's magnetic field in a bandwidth from DC to several kHz \cite{micah,SO,hovde}. In contrast to SQUIDs, which require cryogenic cooling and measure the relative magnetic flux through a pick-up loop, atomic magnetometers operate near room temperature and measure the absolute magnetic field directly by relating it to a frequency and to fundamental physical constants. Currently, the  atomic magnetometer with the highest sensitivity is the spin-exchange relaxation-free (SERF) magnetometer, whose demonstrated sensitivity exceeds  \nolinebreak{$10^{-11}$ G$/\sqrt{\rm Hz}$} (a world record)\cite{record}, with projected fundamental limits below \nolinebreak {$10^{-13}$ G$/\sqrt{\rm Hz}$} \cite{Allred,Kominis,serf,redpaper}. SERF magnetometers also offer the possibility of spatially resolved measurements with sub-millimeter resolution \cite{shah}.

To our knowledge, no one has yet detected the magnetic field from a plant.   Biochemical processes, in the form of ionic flows and time varying ionic distributions, generate electrical currents and time-varying electric fields, both of which produce a magnetic field.   However, contrasted to muscle contraction and brain processes, which have a characteristic time scale shorter than one second, plant bio-processes span several minutes to several days and the expected magnetic field from such processes is correspondingly smaller.  Detection of such small magnetic fields, together with the difficulty of providing the cryogenic support required for SQUIDS, make a sensitive atomic magnetometer a preferred choice.

To mitigate these challenges we turned to a family of plants that exhibit fast bio-processes and thermogenic characteristics while blooming \cite{seymour}. We selected the Titan Arum, or \textit{Amorphophallus titanum}, which is a tuberous plant with the largest known un-branched inflorescence in the world. The inflorescence's single flowers ($\sim$ 500~female and $\sim$~500~male), located at the base of the spadix and enrobed in the spathe, together function as a single plant and flower.  It is indigenous only to the Indonesian tropical forests of Sumatra and grows at the edges of rainforests near open grasslands.   The tuber weighs up to 150~lbs, and grows into a single leaf up to 20 feet tall during the vegetative years.  Reproduction (flowering) may occur every few years after the plant has matured for six years or more \cite{Arum}.

Cultivation of the plant has allowed botanists to study the Titan Arum and its uncommon transformation during the rare blooming years. One of the three most notable characteristics is its size; the tallest recorded bloom occurred at the Stuttgart Zoological and Botanic Garden, Germany, in 2005, and was measured at 2.94m (nine feet, six inches).  The next unusual characteristic is the bloom's distinctive stench of cadaverine and putrescine lasting up to twelve hours after it fully opens, which has given it the name \emph{bunga bangkai} (``corpse-flower'') in Indonesian \cite{springer}.  The smell combined with the spathe's dark purple coloration lure in carrion-eating beetles and flesh-flies that are the putative pollinators.

The third striking feature is the rise and thermoregulation of the spadix temperature, which can reach up to 30$^{\circ}$C above ambient temperature in intervals lasting about 30 minutes over a 12-hour span\cite{torch}.  The heat stimulates the activity of pollinator insects and helps disseminate the scent \cite{web,seymour}.
The Titan Arum's characteristics, including large size and fast biochemical processes, and the availability of a specimen nearing its blooming phase at the University of California Botanical Garden at Berkeley, CA, made it an attractive candidate for this investigation.

\section{\label{estimate}Order of magnitude estimate of expected bio-magnetism}
On a weight-specific basis, plant thermogenesis approaches the rate of heat production exhibited by flying birds and insects; it originates from a large intake of oxygen entering the florets by diffusion\cite{seymour,seymour2}. The Titan Arum has distinct thermal zones extending\nobreak{ }$\gtrsim$\nobreak{ }1\nobreak{ }m upwards from the florets located at the spadix base.   To estimate a possible scale of the plant bio-magnetism, we hypothesize a favorable-case scenario (from the point of view of generation of a magnetic field), modeled by a bi-directional ionic transport of oxidation/reduction chemical reactants.  We approximate this ionic transport by two long parallel wires located at the core of the spadix and separated by a distance $d = 10\ \rm\mu$m (a characteristic plant cell size).

The work required to raise the temperature of a characteristic mass $m=1$\nobreak{ }kg of plant material (mostly water) by $\Delta T \approx 10^{\circ}$C above the ambient environment is:
\begin{equation}
W=\Delta T mc \simeq 42~\rm kJ,
\end {equation}
where $c$ = 4.2 kJ/kg is the specific heat of water.  In a characteristic thermogenic time interval of $t\sim 30$ min this corresponds to a power of:
\begin{equation}
P=\frac{W}{t}\approx 20~\rm W,
\end {equation}
which is commensurate with the calorimetry measurements performed with other thermogenic plants\cite{calorimetry}.

Assuming 1 eV per oxidation event\cite{life}, the magnetic field induced by the bi-directional currents at the nearest gradiometer sensor, positioned at a distance $D$ = 20 cm from the plant core, is:
\begin{equation}
B\propto\frac{Pd}{D^2},
\label{10uG}
\end {equation}
which leads to an expected magnetic field magnitude on the order of $\rm 30 \ \mu G$.

The magnetic field variations due to bio-magnetic processes are expected to occur on a time scale ranging from 15 to 30 minutes; the output of the magnetometer can therefore be averaged over one minute intervals. This would give a sensitivity better than 100~nG per point using the atomic magnetometer (described in section \ref{setup}), which is more than sufficient to resolve the magnetic field in this scenario.
\begin{figure}
  \includegraphics[width=2.5in]{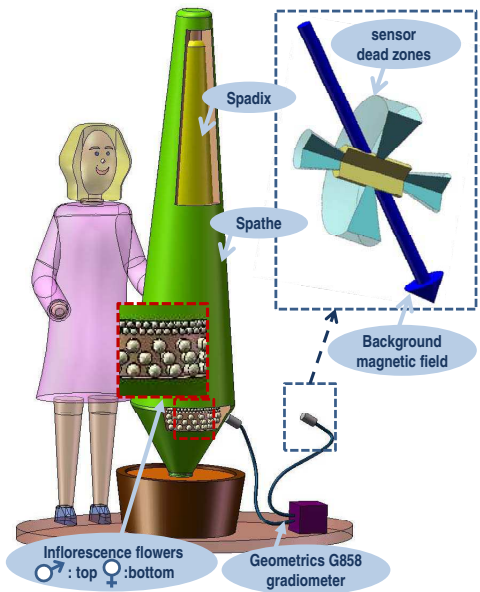}
  \caption{(Color on line) Experimental setup: The Geometrics G858 atomic gradiometer is positioned with one sensor near the spathe where pollination occurs. The other sensor is used to subtract the ambient magnetic field noise. [Insert: Each sensor's dead sensing zones (light blue) lie within 30$\rm^{\circ}$ of the sensors axis and within 30$\rm^{\circ}$ of the plane perpendicular to it. The downward pointing arrow indicates the direction of the ambient magnetic field.]  The sensor axes are parallel and $\rm\sim 45^{\circ}$ to the ambient magnetic field, which is inclined $\rm\sim60^{\circ}$ to the vertical and commensurate to the local earth magnetic field \cite{earthfield}.}
\label{fig:flower+mag}
\end{figure}

\section{\label{setup} Experimental set up and Environment}
The plant chosen for the experiment, nicknamed ``Trudy'' $\rm (Fig.~\ref{fig:titan})$, was blooming for the second time at fourteen years of age, reached a peak height of $\sim$~2~m, and was kept in a heated greenhouse approximately $\rm8\times8\times8 \ m^3$ in size. The experimental environment includes four main types of magnetic-field noise, each one being on a different time scale.  The San Francisco Bay Area Rapid Transit electric-train system (BART) causes fluctuations in the magnetic field on the order of 0.5 mG on a time scale ranging from a fraction of a second to a minute; those fluctuations are absent from $\sim$1~AM to 5 AM when BART suspends operation.  Visitors, during the garden opening hours (9~AM~-~5~PM), cause magnetic field fluctuations on a several second to a minute time scale. Sudden displacement of the plant pot and/or the magnetic sensors add stepwise changes in the magnetic field and gradients.  Another intermittent magnetic field noise is caused by the greenhouse temperature regulation mechanism which includes two electric heaters and two large fans located near the ceiling of the greenhouse; a thermostat turns on the heaters and fans every 15 to 30~minutes maintaining a temperature ranging from 25$\rm^{\circ}$C to 30$\rm^{\circ}$C throughout the greenhouse. This causes corresponding sudden spikes and stepwise magnetic field and gradient variations.

The experimental setup is shown in $\rm Fig.~\ref{fig:flower+mag}$.   A commercial G858 Geometrics cesium atomic magnetometer/gradiometer was selected for the experiment.  The G858 is a scalar (as opposed to a vector) sensor, and measures the projection of the magnetic field onto the prevailing field axis\cite{magbook,geomag}.  The G858 has a sensitivity of 100 nG (at 1 second cycle rate), a temperature dependence of 500nG/C$^\circ$\cite{G858}, and an operating principle derived from the techniques pioneered by Bell and Bloom\cite{Bloom,BellBloom}.

One sensor was positioned $\sim$~5~cm from the spathe near the location where pollination and thermogenesis occur and where we speculate the plant bio-magnetic activity may be strongest.  The other sensor was positioned $\sim$ 0.5 m from the plant, served to subtract the ambient magnetic field.
A static magnetic field gradient throughout the greenhouse was measured to be approximately 10 $\rm\mu G/cm$ and added a constant offset between the outputs of the two magnetometer sensors, which depended on the positioning of the sensors in relation to the gradient direction.  The sensor axes were aligned to have the ambient magnetic field direction fall outside the magnetometer dead zones (which lie within 30 degrees of the sensor axis and within 30 degrees of the plane perpendicular to it).
  \begin{figure}
  \includegraphics[width=3.5in]{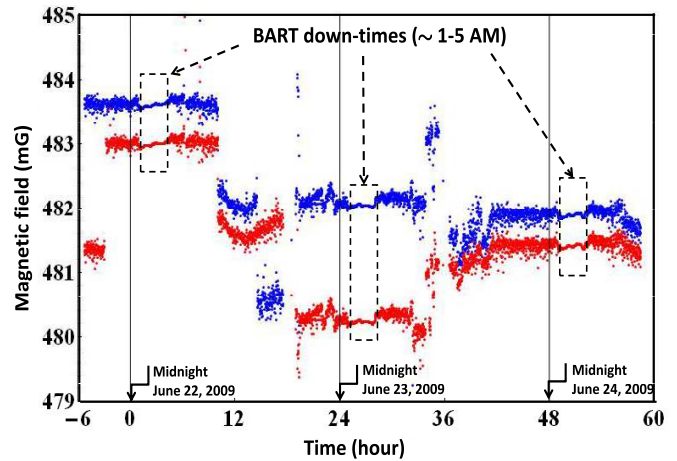}
  \centering
  \caption
{(Color on line) Data from the two magnetometer channels spanning 3 days.  The local earth magnetic field is $\sim$500 mG.   The three rectangular boxes indicate the magnetically quiet periods when the BART operations are suspended from $\sim$1-5~AM.   Discontinuities in the data correspond to shifting of the plant and/or the magnetometer sensor heads.   Large magnetic-field fluctuations are seen during the U.C. Botanical Garden open hours (9~AM - 5~PM).  The difference between the two magnetometer channels depends on their position relative to the ambient magnetic field gradients.}
\label{fig:3days}
\end{figure}

\begin{figure}
\center
  \includegraphics[width=3.5in]{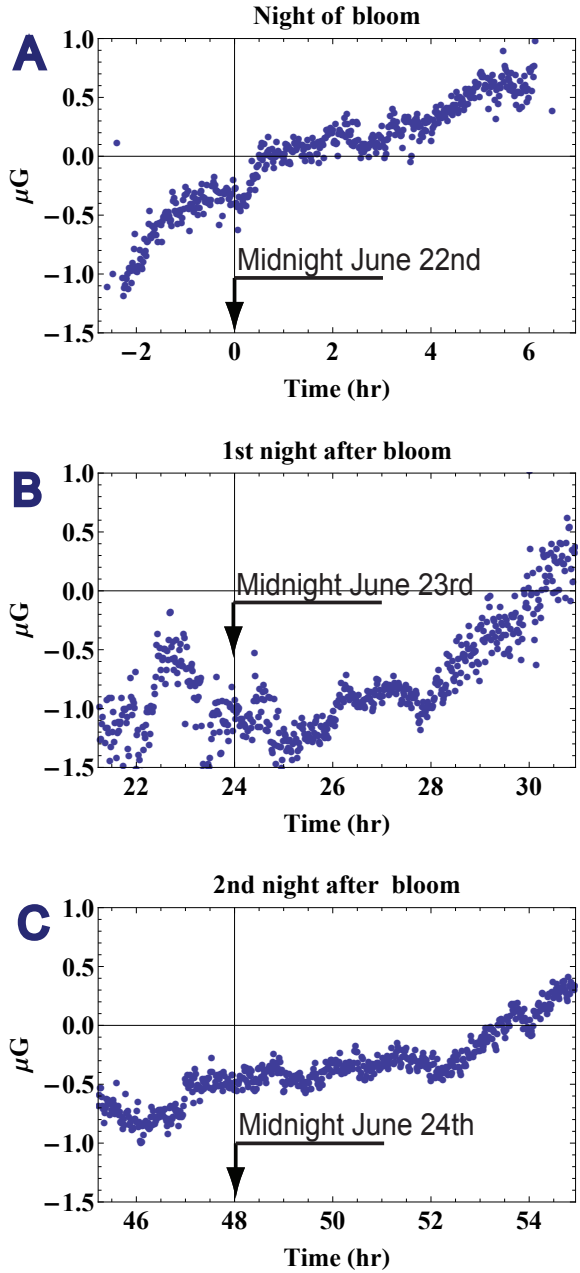}
  \caption{(Color on line) Gradiometer signal (difference magnetic field). (A): 9 hr segment: night of bloom. (B\&C): for comparison, same 9 hr segments on the following two nights.  Data averaged over one minute intervals.  We attribute the overall magnetic field increase to a gradual change in temperature and corresponding residual temperature dependence of the sensor ($\rm0.5~\mu G/^{\circ}C$). The time scale of that drift is significantly longer then the time scale we expect from the plant biomagnetic activity.  On each of the three nights, the magnetic field noise at 1 mHz, in a 0.5 mHz bandwidth (corresponding to events lasting $\sim$10-30 min), is $\approx$ 0.6 $\mu$G$_{\rm rms}$.}
\label{fig:3nights}
\end{figure}
\begin{figure}
\center
  \includegraphics[width=3.3in]{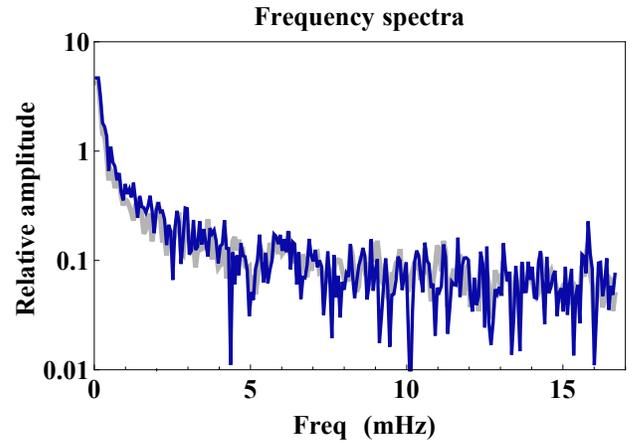}
\caption{(Color on line) Frequency spectra of data shown in~Fig.~\ref{fig:3nights}.  Dark (blue):~night of bloom (Fig. 4A). Light (gray): average of subsequent two nights (Figs. B-C).  A biomagnetic event of a magnitude larger than 0.6~$\mu$G and lasting 10-30 minutes would appear as a feature between 0.5~and~1.5~mHz rising above the overall 1/f noise. }
\label{fig:fft}
\end{figure}

\section{\label{results} Results}
Figure \ref{fig:3days} shows the outputs from the two gradiometer channels.  Data were collected over a period of three consecutive days starting on the evening of June 22, 2009. We visually observed the anthesis (beginning of the blooming phase) at approximately 9 PM on the night of June 22.   Midnight on that night is zero on the the time axis.  Discontinuities in the data were caused by inadvertent moving of the pot and/or the sensors.  The gaps in the data occurred during data downloading and apparatus maintenance.  The BART-free time periods ($\rm\sim$1 - 5 AM) are clearly visible as relatively magnetically quiet periods on each of the two magnetometer channels.  Large magnetic field fluctuations are also visible during the Garden open hours (9 AM - 5 PM).  Figure \ref{fig:3nights} shows the difference magnetic field, as measured by the two sensors.  Three segments of data are shown, from 9 PM to 6 AM, on three consecutive nights, starting on the night of the bloom.  The power spectrum of the first segment is shown in Fig. \ref{fig:fft}.   The amplitude of the magnetic field noise at 1 mHz in a 0.5 mHz bandwidth (frequency range equivalent for events lasting from 10 to 30 minutes) is 0.6 $\rm \mu$G, and was was similar for all three nocturnal time segments.

\section{\label{conclusion} Discussion, Conclusion, and Outlook}
The above result sets an upper bound of 0.6 $\mu$G on the amplitude of bio-magnetism from the plant while blooming, projected onto the ambient magnetic field direction, for events in the 10 to 30-minute cycle range ($\sim$1 mHz), and at a distance of $\sim$ 5 cm from the surface of the spathe.
Within the framework of the simple model of section~\ref{estimate}, this is significantly lower than the expected magnetic field.  However the plant bi-directional ionic currents in the model may instead be distributed in a more complex geometry than the two parallel wire model used in this analysis and with correspondingly more magnetic field cancellation.  In a limiting case, there is no net magnetic field if the ionic flow is modeled by a core current enclosed by a cylindrically distributed, opposite, and counter-propagating current.  The Titan Arum spadix does not have perfect cylindrical symmetry and one may expect a departure from total cancellation of the magnetic field.
\newline

To further investigate plant biomagnetism greater magnetic field detection sensitivity is necessary.   Several options are possible: using an array of micro sensors to better locate and resolve the source of the magnetic field and to more effectively subtract the fluctuations and drift of the ambient magnetic field and its gradients; moving to a more isolated environment that is removed from public access and electrical devices, magnetically shielding the plant to eliminate the fluctuations of the magnetic field and gradients, and/or selecting a smaller plant with fast bio-processes like the Sensitive Plant \textit{(Mimosa pudica)} or the Sacred Lotus \textit{(Nelumbo nucifera)}. A smaller plant size would facilitate the complete coverage of the thermogenic zones.  Concurrently measuring the spatial distribution and the variations of the plant temperature with an infrared camera and correlating that measurement to the magnetic field measurement would correspondingly yield a better sensitivity.

\section {Acknowledgment}
The authors are indebted to the University of California Botanical Garden staff who generously granted after-hours access to the garden facilities.  We also acknowledge stimulating discussions with Robert Dudley from the department of Integrative Biology, with Lewis Feldman, Steve Ruzin, and Peggy Lemaux from the department of Plant and Microbial Biology, and with Philip Stark from the Department of Statistics, at University of California at Berkeley. Invigorating exchanges with our colleague Todor Karaulanov helped the understanding of the processes at hand.   This project was funded by an ONR MURI fund and by the U.S. Department of Energy through the LBNL Nuclear Science Division (Contract No. DE-AC03-76SF00098).

\bibliographystyle{apsrev}

%

\end{document}